\documentclass[referee]{aa} % for a referee version
\usepackage{hyperref}
\usepackage{graphicx}
\usepackage{txfonts}
\usepackage{threeparttable}
\usepackage{subcaption}
\usepackage{flushend}
\begin{document}

   \title{Responses of the X-ray spectrometer/imager STIX onboard Solar Orbiter}

   \subtitle{}

   \author{Hualin Xiao
          \inst{1,2}
           \and Olivier Limousin \inst{3}
          \and Ewan Dickson \inst{4}
          \and S\"am Krucker\inst{1} 
         }

   \institute{University of Applied Sciences and Arts Northwestern Switzerland (FHNW), 5200 Windisch, Switzerland \\
              \email{hualin.xiao@fhnw.ch}
           \and   The Physical Meteorological Observatory and World Radiation Center (PMOD/WRC) , 7260 Davos, Switzerland
         \and IRFU, CEA, Université Paris-Saclay and Université Paris Diderot, AIM, Sorbonne Paris Cité, CEA, CNRS, 91191 Gif-sur-Yvette,
         France
             }

   \date{\today}

  \abstract
   { 
Solar flares are explosive events that release X-rays from hot plasma and accelerated electrons. The STIX instrument on the Solar Orbiter provides imaging spectroscopy of solar X-ray emissions from 4 to 150 keV. 
   } %leave it empty if necessary  
  % {context.}
  % aims heading (mandatory)
   {
   To interpret the STIX data accurately, understanding the instrument's response is crucial. 
   }
   {
   Given the complexity of interactions of X-rays with the instrument, we developed a detailed Monte Carlo model for STIX based on Geant4. The model accurately depicts the instrument's components, such as grids, detectors, X-ray windows, and collimators, with their responses.
    }
   {We studied various effects, including grid shadowing, fluorescent X-rays emitted by materials in STIX, and grid transmission, to assess their impacts on STIX's scientific goals. Model validation was performed using Crab Nebula observations, a standard calibration source that provides reliable ground truth for X-ray instruments. Our simulations align with the Crab Nebula observations within the uncertainties, thereby validating the accuracy of the Geant4 model and showcasing its potential for interpreting STIX data. With the help of the generated response matrices, which are indispensable for solar spectroscopy, we discuss the applications and limitations of the model for future STIX data analysis.
  }
 {}
  \titlerunning{STIX response matrices}
  \authorrunning{Hualin Xiao and the STIX team}
   \maketitle
%-------------------------------------------------------------------

\section{Introduction}
 Solar flares are sudden releases of magnetic energy in the solar atmosphere that produce intense radiation across the electromagnetic spectrum, from radio waves to gamma rays. 
 Solar Orbiter is a space mission of international collaboration between ESA and NASA that aims to study the Sun and its heliosphere from close distances and high latitudes (\citep{SolarOrbiter2020}). One of the scientific objectives of Solar Orbiter is to understand how solar flares and coronal mass ejections (CMEs) are triggered and how they affect the solar wind and interplanetary space. 
 X-rays are one of the most direct diagnostics of solar flares, as they are emitted by both thermal plasma heated to millions of degrees and non-thermal electrons accelerated to relativistic speeds.
The Spectrometer/Telescope for Imaging X-rays (STIX) is one of the ten instruments on board Solar Orbiter which is dedicated to observing solar hard X-ray emissions from 4 to 150 keV (\citep{stix2020}, \citep{stixdatacenter}). STIX uses a Fourier-transform imaging technique based on a set of 32 sub-collimators, each consisting of a pair of tungsten grids separated by 55 cm \citep{stix2020}. 
The grids modulate the incoming X-rays and create Moir\'{e} patterns on 32 pixelated Cadmium-Telluride (CdTe) detectors behind them. Images of solar flare X-ray sources can be reconstructed from the relative count rates of the detectors by applying an inverse Fourier transform on the ground. STIX also provides spectroscopic information by measuring the energy spectrum of the detected X-rays.

X-rays from the solar flares can be absorbed, scattered, or create secondary X-rays when they interact with materials in STIX before being detected by the X-ray detectors. On the other hand, the deposition energy recorded by the detector may differ from that of the incident X-rays. This is because the energy of the X-rays can be scattered upon entering the detector, or it can be used to produce characteristic X-rays that can escape from the detector. Additionally, the detector itself is susceptible to bulk effects, near-surface effects, and other influences \citep{oliver}. These factors can have implications for imaging and spectroscopy. To properly interpret STIX observation data, it is crucial to have a comprehensive understanding of how the instrument responds to X-rays.

In this paper, we introduce a Monte Carlo simulation model developed based on Geant4 \citep{geant4, geant4recent} for STIX. This model allows the simulation of X-ray interactions with various materials and detector responses. The model incorporates geometry and material information for the individual components, as well as the modelling of detector responses and the behaviour of readout electronics, based on our current understanding. We investigate the impact of instrument effects on observations, such as grid shadowing, grid transmission, and fluorescent photons, which have not been taken into account in the current IDL spectroscopy and imaging software. 
Model validation is performed using Crab Nebula observations, a persistent astronomical source that provides reliable ground truth for calibrating X-ray instrument responses. Our findings indicate that the simulation results align with the Crab Nebula observations within the uncertainties, thereby validating the model's accuracy. Furthermore, we present response matrices generated using the model, along with corresponding validations. Finally, we provide a discussion of the model's applications and limitations for future STIX data analysis.

\section{Instrument mass model and the simulation package}

Geant4 is a software package developed at CERN that allows for modelling the interactions of energetic particles and photons with matter.
It utilizes the Monte Carlo method to accurately simulate the stochastic nature of these interactions. 
Geant4 is written in C++ and has found extensive applications in particle physics, medical physics, and space projects\cite{geant4}.
A typical Geant4-based user application consists of modules to define the geometry and material of the mechanical parts, relevant physics processes, particle tracking methods, and the primary particles.  In the following sections, we will introduce the above modules developed for STIX, as well as runs conducted to validate the model.
\subsection{Instrument geometry modelling and materials}
There are two primary approaches to representing geometry in Geant4: Constructive Solid Geometry (CSG) and the tessellated model. CSG is a technique to create complex geometric shapes by combining simple primitives (such as boxes, spheres,  cylinders, and polyhedrons) using boolean operations, such as union, intersection, and difference;  while the tessellation method employs triangles to represent the surfaces of the geometry.  Tessellated geometries are usually used when the geometry shapes are complex, and they can be converted from CAD models. As such, this may result in decreased accuracy during the conversion, particularly for smaller components. Volumes modelled using the CSG method allow for faster simulations and usually have a more accurate representation of objects with simple shapes, but it requires a lot of effort to model complex geometries manually.   As such, we used both methods for this work.

\begin{figure*}[htbp]
    \centering
    \begin{subfigure}[c]{0.32\textwidth}{
        \centering
        \includegraphics[width=0.7\linewidth]{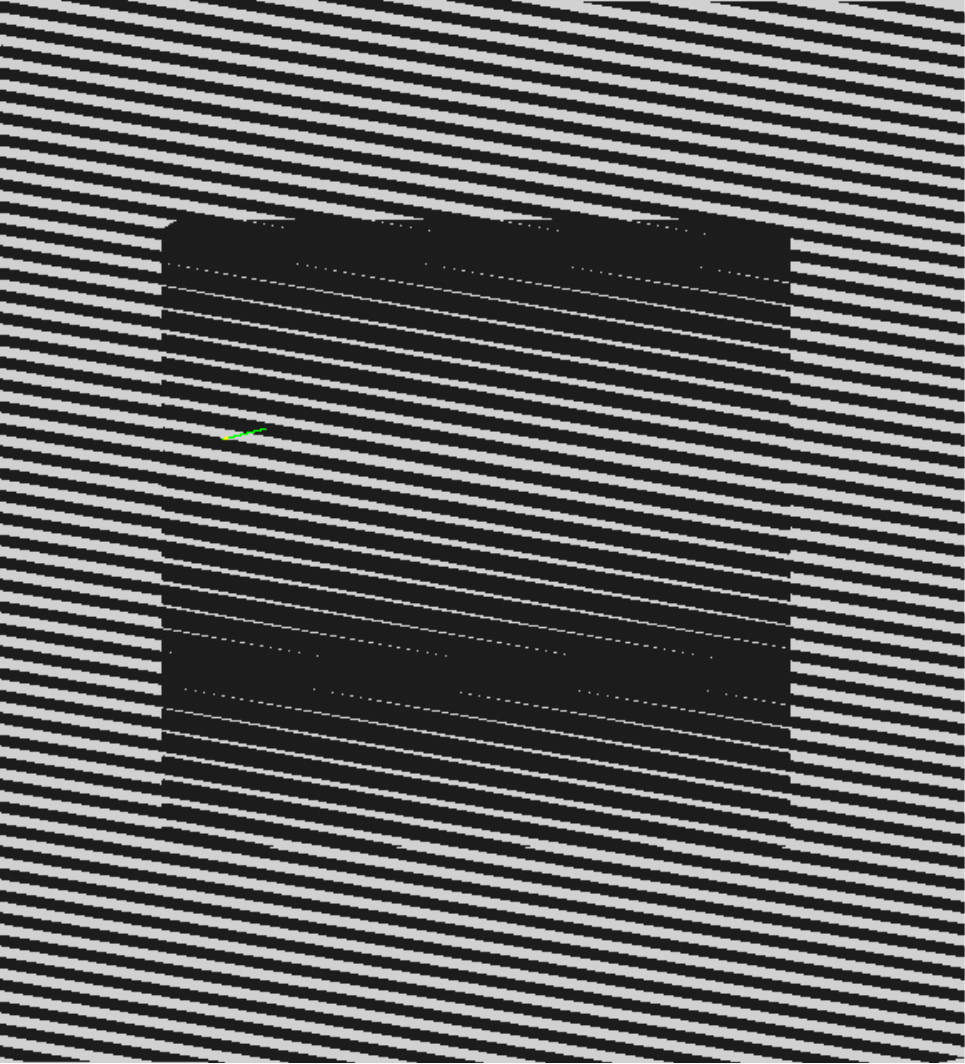}
        \label{fig:grids}
       % \caption{Collimator \#20.}
        }
    \end{subfigure}
    \begin{subfigure}[c]{0.32\textwidth}{
        \centering
        \includegraphics[width=0.75\linewidth, angle=90]{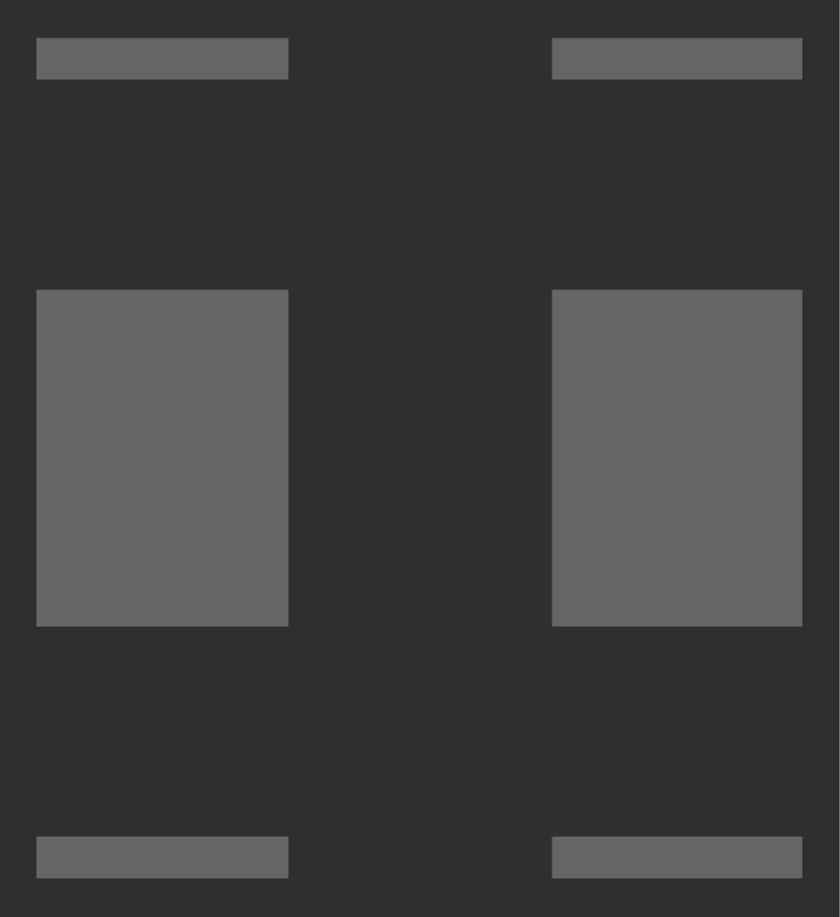}
        \label{fig:cfl}
       % \caption{CFL collimator.}
        }
          \end{subfigure}
    \begin{subfigure}[c]{0.32\textwidth}{
        \centering
        \includegraphics[width=0.75\linewidth,angle=90]{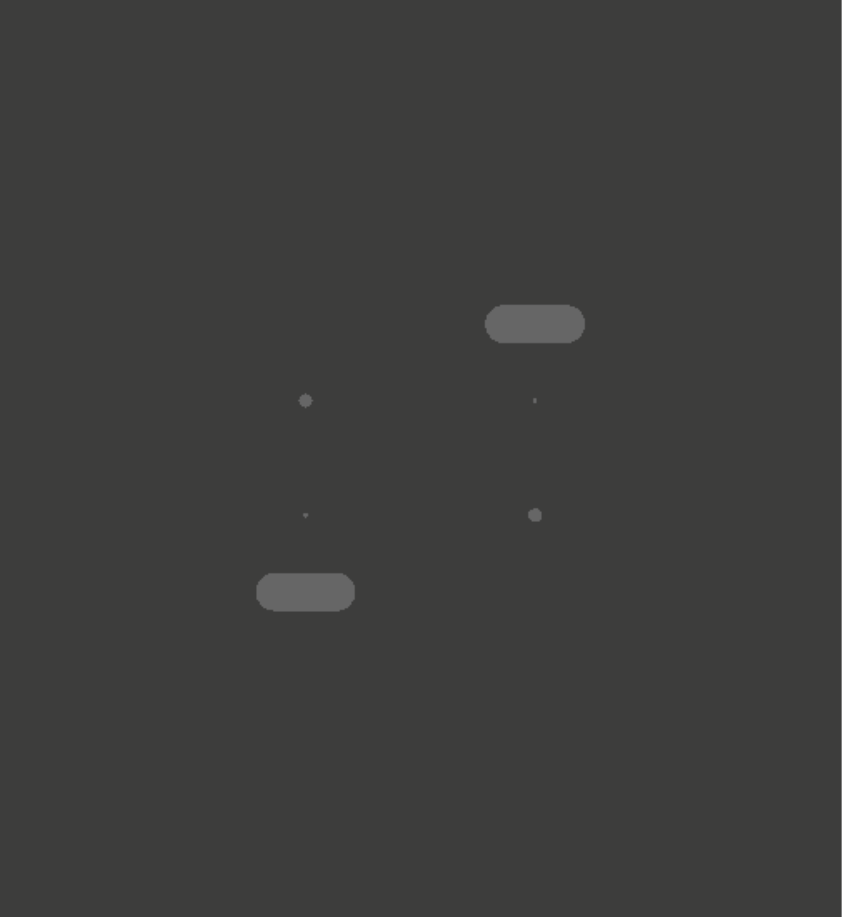}
        \label{fig:bkg}
      %  \caption{BKG collimator.}
        }
    \end{subfigure}
    \caption{View of constructed front and grids for collimator \#20 (left) and two special STIX collimators: CFL (middle) and BKG (right) with Geant4. They are modelled using Constructive Solid Geometry (CSG) primitives. The front and rear grids are overlaid. The Moiré pattern can be seen. The grey regions in the CFL and BKG figures are apertures.}
    \label{fig:collimators}
\end{figure*}

In the STIX instrument, 30 pairs of grids are utilized to measure spatial Fourier components of X-ray sources, with two special collimators for flare localization and background measurement;
each grid pair 
comprises a rear grid and a front grid, consisting of evenly spaced tungsten strips.  The total number of tungsten strips reached 5,800. We modelled the STIX grids using CSG primitives in order to reduce computing resources during simulations.
In the simulation package, each grid strip is described as a 0.4 mm thick parallelepiped using CSG primitives. The dimensions and separation between neighbouring strips, as well as the rotation angle, 
are determined through ground calibrations. Subsequently, the grid strips are assembled and placed within a frame. The left panel of Fig. ~\ref{fig:collimators} shows a plot overlaying the constructed grid and rears for sub-collimator \#20, visualized with Geant4. 
A moiré pattern is visible in the figure. The two special collimators are also modelled using CSG primitives. Their representations in Geant4 are shown in the middle and right panels of Fig. ~\ref{fig:collimators}.  
\begin{figure}
        \centering
        \includegraphics[width=0.4\linewidth, angle=90]{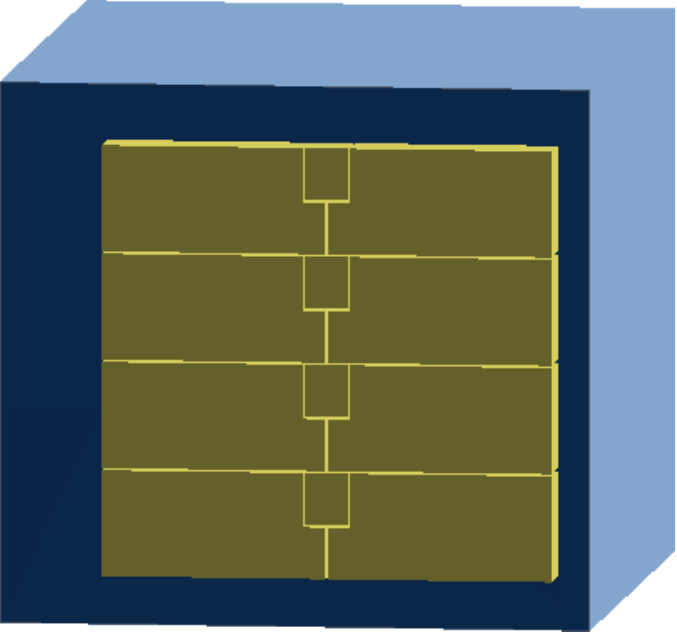}
        \caption{View of constructed STIX X-ray detector in Geant4.}
    \label{fig:cdte}
\end{figure}
STIX uses 32 identical pixelated  CdTe detector modules to detect X-rays. They are also modelled using the CSG method in the package. 
Fig. \ref{fig:cdte} shows the view of a detector module in Geant4.  Starting from the detection surface and going downward, there are platinum coatings, 1 mm thick CdTe (divided into 12 pixels), and the CdTe bottom coating layers. Under the CdTe are metal electrodes and a metal-coated resin box. The ASIC inside the resin box is approximated using a silicon plate.  
\begin{table*}
\caption{Summary of modelling methods and materials for key components.  
The tessellated models were converted from those in STIX CAD files. }
\label{tab:stix-instrument}
\begin{tabular}{ccc}
\hline
Part & Geometry model type& Material \\
\hline
Front X-ray window & Tessellated & Beryllium (Type S-200F) with solar black coating \\
Rear X-ray window & Tessellated &  Beryllium (S-200F) \\
Fine grid cover & Tessellated& Kapton \\
Attenuator & Tessellation & Aluminum 7075 \\
Grids & CSG& Tungsten \\
DEM entrance &  CSG & Multi-layer insulation \\
X-ray detectors &CGS & CdTe with coating layers\\
\hline
\end{tabular}
\end{table*}
The shapes of certain components, such as the front and rear windows of the X-ray, the grid supports, the aluminum attenuator, the multi-layer insulator, the cold plate, the entrance window, and the enclosure for the Detector Electronics Module, are obtained from CAD models using G4CAD. This tool tessellated 3D models in the CAD files and wrote the geometry description into the GDML format supported by Geant4.  The cables, screws, and the grid supports, which are  located either away from the CdTe detectors or the x-ray path, are excluded to reduce the computational cost since their impacts on X-ray detection are negligible.
The materials for each component are defined according to our best knowledge. 
Table \ref{tab:stix-instrument} presents the material information of the key components included in the package. 
The composition details for alloys were taken from the manufacturer's datasheets.

\begin{figure*}[htbp]
    \centering
    \includegraphics[width=0.9\linewidth]{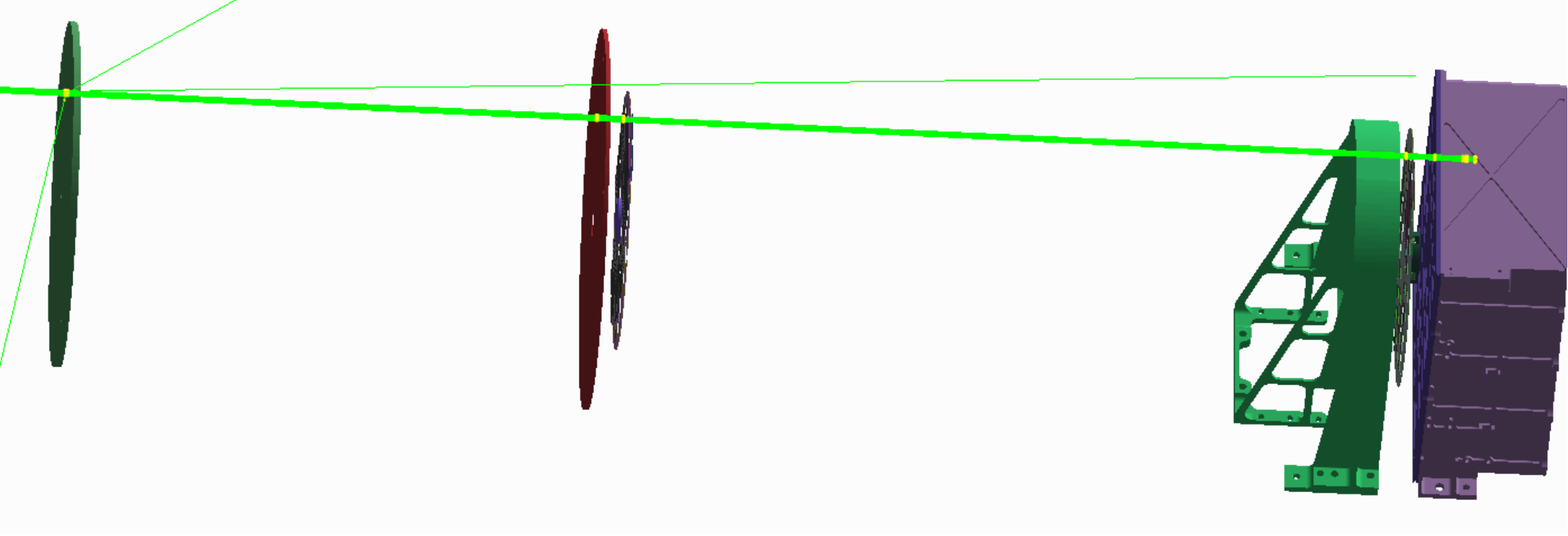}
    \caption{View of the STIX instrument module in Geant4,
    showcasing the trajectories of 20 keV X-rays during a test run. 
    Tungsten grids and X-ray detectors are accurately
    represented using Constructive solid geometry (CSG), 
    while the remaining components are tessellated from the STIX CAD file. To speed up simulations, the instrument model excludes non-essential components such as cables and screws located behind the detector units, as they have negligible contributions to X-ray detection. X-rays are directed towards the X-ray window, with the majority being absorbed by the X-ray detector.}
    \label{fig:stixfull}
\end{figure*}

Fig.~\ref{fig:stixfull} shows a visualization of STIX instrument model in Geant4,  
showcasing the trajectories of 20 keV X-rays during a test run.

\subsection{Physical processes}
Physical processes have to be defined by users in Geant4 based simulation packages.
Given our focus on low-energy physical processes, the Livermore low-energy electromagnetic models were employed. 
The models are capable of describing the interactions of electrons and photons with matter down to about 250 eV, a proximity to the K-shell Auger peak from carbon, using interpolated data tables based on the Livermore library.
The Livermore models cover essential processes for gamma rays, including the photoelectric effect, Compton scattering, Rayleigh scattering, and conversion. For electrons, ionisation, and Bremsstrahlung are included.  In addition to the processes in the Livermore models, 
the atomic de-excitation processes fluorescence and Auger electron emission are  activated for processes producing vacancies in atomic shells.

\subsection{Detector response modelling}
\label{sec:effects}
\subsubsection{Charge collection}
When a photon deposits energy $\Delta E$ at a certain depth $x$, which is measured from the cathode,  in the CdTe detector, the average number of  created electron-hole (e/h) pairs is given by $N_{\rm e/h}=\Delta E/E_{\rm pair}$. $E_{\rm pair}$ is the energy to create an e/h pair.  In the case of CdTe, $E_{\rm pair}=4.43 $eV \citep{raddetbook}.  The created charge can subsequently drift to the opposite electrons under the action of the external electric field.  Due to the finite lifetime of the charge carriers, only a fraction of the charge can be collected. The charge collection efficiency $f_{\rm Hecht}(x)$  in a can be expressed by  the Hecht relation:
\begin{equation}
\label{eq:hecht}
f_{\rm Hecht}(x)= \frac{\lambda_{\mathrm{e}}}{d} \left(1 - \exp\left( \frac{x - d}{\lambda_\mathrm{e} } \right)\right) + \frac{\lambda_\mathrm{h}}{d} \left(1 - \exp\left(-\frac{x}{\lambda_\mathrm{h}}\right)\right).
\end{equation}
Here,$d$ represents the detector thickness, and $\lambda_\mathrm{e}$ and $\lambda_\mathrm{h}$ denote the drift lengths of electrons and holes, respectively. The drift length $\lambda_{\mathrm{e/h}}$ is given by $\mu_{\mathrm{e/h}} \tau_{\mathrm{e/h}} E$, where $E$ represents the electric field strength, and $\tau$ the lifetime of the carriers.  The default values of parameters 
are given in the Table \ref{tab:hechtparam}.  
\begin{table}[h]
\centering
\caption{Values of parameters for the Hecht formula. The values for $\mu_{\mathrm{e/h}}$, $ \tau_{\mathrm{e/h}} $ are taken from \citep{oliver}. Most of the time, STIX has been operated with a high voltage of 300 V. }
\label{carrier_table}
\begin{tabular}{cccc}
\hline
Carrier & $\mu_{\mathrm{e/h}}$ (cm$^2$/Vs) & $ \tau_{\mathrm{e/h}} $ (us) & $\lambda$ (mm)  at 300 V/mm \\ \hline
Electron & 1100 & 3 & 99 \\ \hline
Hole & 100 & 2 &  6 \\ \hline
\end{tabular}
\label{tab:hechtparam}
\end{table}
As described in \citep{oliver,nearsurface}, the collection efficiency for the charge created near the detector surface is lower due to the near-surface damage in the crystal. This effect can be modelled by multiplying the created charge by a factor $f_{\rm NS}(x)=1-R_0\cdot \exp(-x/L)$. 
In the package, the collected charge is computed for each "step," which is a discrete particle movement as it passes through a medium in Geant4, utilizing the previously mentioned equations. Therefore, the collected charge $Q_{\rm sum} = N_{\rm e} e$ for each pixel and each event is \begin{equation}
Q_{\rm sum} = N_{\rm sum} e = \frac{e}{E_{\rm pair}}\sum _i \Delta E_i f_{\rm Hecht}(x_i) \cdot f_{\rm NS}(x_i),
\label{eq:qsum}
\end{equation}
where $N_{\rm sum}$ and $e$ denote the number of the collected electrons at the electrode and the electron charge, and $x_i$ the distance from the location of the “step” to the cathode, respectively.  The second term, denoted as $E_{rm vis}$,  is  the visible  deposited energy. 

\subsubsection{Energy resolution smearing}
The recorded energy can deviate from the $E_{\rm vis}$ due to the energy resolution smearing. 
The energy resolution ($\sigma_E$) can be derived from a combination of factors, 
including the Fano factor, $F$, and the influence of electronic noise, denoted as $\sigma_{\rm noise}$. 
It can be given by:
\begin{equation}
\sigma_E = \sqrt{(F \cdot N_{\rm sum})^2 + \sigma_{\rm noise}^2}
\end{equation}
In the package, we use F= 0.15 \citep{raddetbook}, along with an experimentally measured $\sigma_{\rm noise}$ of 0.46 keV.
In order to account for the energy smearing effect, the visible energy, denoted as $E_{vis}^{'}$, is  sampled from the Gaussian distribution with $\mu=E_{\rm vis}$ and $\sigma = \sigma_{\rm E}$.
$E_{vis}^{'}$ is the expected measured energy. 
It's worth noting that the actual noise level of 
each pixel varies slightly and also changes 
because of detector degradation.

\subsection{Event recording}
The pixel summed deposited energy, without considering detector effects,  
the collected charge and visible energy and triggering status,  
as well as the information of particles entering the CdTe detector 
surface, are recorded on an event-by-event basis.
Additionally, the information of the initial particle, 
such its position, direction, particle type, and energy, is also included in the records.

\section{Model verification with the Crab Nebula }
STIX recorded tens of thousands of solar flares each year, but due to the lack of ground truth of their energy spectra, it is challenging to validate our model using the flares themselves. 
Cross-calibration with other near-earth instruments  such as HXI and Fermi are also challenging, due to the different energy ranges and energy ranges of instruments as well as the viewing angle effects.

The Crab Nebula is one of the brightest persistent X-ray sources in the sky, considered as the standard candle for calibrating X-ray responses of various space instruments (see, e.g., \citep{crabcali1,crabcali2}). 
The Crab Nebula was in the field-of-view of all STIX pixels 
from 2020-05-24T00:30:00Z to 2020-05-24T01:50:00Z. 
STIX was in observation mode and the sun was quiet during the period. 

The photons from Crab Nebula pass the same materials as those from solar flares before being detected by the X-ray detectors; therefore, the collected data is ideal for verification of the models in the simulation package.

Three simulation runs for different Crab Nebula locations were conducted in order to validate the computer model. In the simulation runs, primary photons were randomly generated within a 12 cm circular 
area in front of the X-ray entrance window, with directions parallel to the vectors from the Crab Nebula at the start, center, and end locations during the selected time range to STIX, respectively, and their energies were sampled from the Crab Nebula spectrum reported 
in Refs. $f(E)$ \citep{crab1,crab2} in the
energy range from 3 keV to 150 keV: 
\begin{equation}
    f(E)=9.59E^{-2.108},
    \label{eq:crab}
\end{equation} 
where $f(E)$ is in units of photons cm$^{-2}$ s$^{-1}$ keV$^{-1}$.
All the instrument effects described
in chapter \ref{sec:effects}  were included
during the simulation. 
 \begin{figure}[ht]
        \centering
          \includegraphics[width=0.7\linewidth]{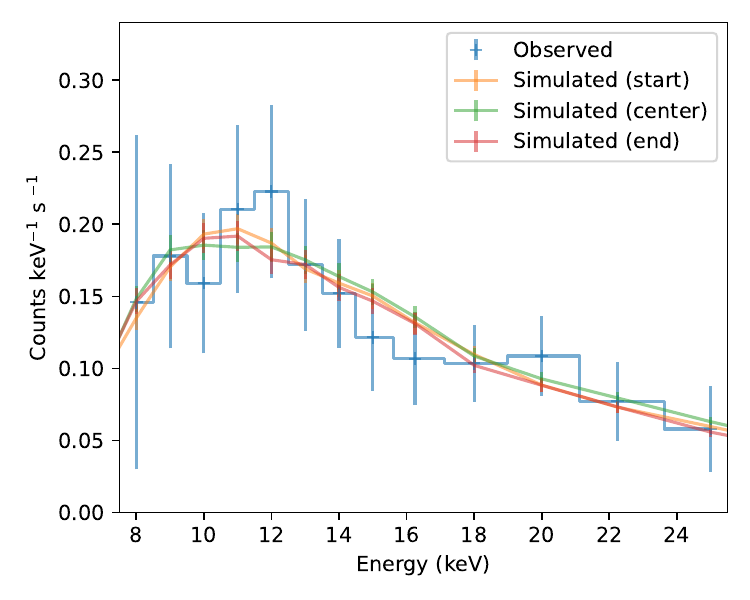}
              \caption{
  STIX background-subtracted count rate spectrum
  observed for the Crab Nebula from 2020-05-24T00:40:00Z to 2020-05-24T01:50:00Z, 
  and Monte Carlo simulated spectra for the Crab Nebula at the start, center and end locations within the selected time range.  
  The counts recorded by the big pixels of the thirty imaging detectors are included in the spectra.
  The simulated energy spectrum has been
  smeared with the instrument's energy resolution. }
    \label{fig:crab}
\end{figure}
%Crab number paper https://iopscience.iop.org/article/10.1088/0004-637X/801/1/66/pdf
%background time range
%error bar
%make a plot for bkg spectrum
%extend sources

The Level-1 dataset \citep{stixdatacenter} for the time range 2020-05-24T00:30:00Z to 2020-05-24T01:40:00Z was selected as the signal data. It is worthwhile mentioning that the counts in energy bins above 26 keV were not down-linked due to the constraint on the telemetry rate during the commissioning phase.  
For background subtraction, we selected the dataset recorded from 2020-05-23T16:20:00Z to 2020-05-23T17:30:00Z, when the Crab Nebula was fully obscured by the Sun and the temperatures of detectors followed the same trends as those in the selected signal time range. The dead time was subtracted before the background subtraction, and only counts recorded by the big pixels were used for this analysis.  

Fig.~\ref{fig:crab} shows the simulated count rate spectra as well as the background-subtracted count rate spectrum measured by STIX.  
 As can be seen in the figure, the two spectra agree within a 1-sigma uncertainty.
 
It is worthwhile to mention that it is challenging to use the observation data for solar flares for this purpose, as the true spectra of solar flares are unknown.
%To determine There are data collected by other instruments, such as Fermi and HXI \cite{fermi, hxi};  however, the instruments have different energy ranges and there may be 

\section{Study of  instrumental effects using the Monte Carlo methods}
STIX data are susceptible to a variety of instrumental effects.
Understanding and quantifying these effects is essential for accurately
interpreting the measured spectra and reconstructing flare images. 
Monte Carlo simulations allow us to investigate instrumental effects in detail. 
This chapter will study the key instrumental factors
that can impact the measured X-ray spectra and imaging in STIX.

\subsection{Tungsten grids}

STIX's imaging principle is based on Moiré patterns created by 
thirty pairs of front and rear tungsten grids. 
An ideal grid would have zero thickness and its strips are completely 
opaque to photons with energies of 4 -- 150 keV. However, achieving this level of perfection is not feasible in reality.  The strips in STIX grids have a thickness of 400 $\mu$m. 
This specific thickness was chosen after considering various factors, 
including manufacturing challenges, mechanical strength, and transmission requirements.
%Photons either pass through the slits or interact with the strips in the front, rear or both grids before they reach the X-ray detectors. 

Photons up to tens of keV can be fully absorbed, i.e., opaque, by tungsten strips of such thickness; however, photons at higher energies can pass through, or undergo scattering or create fluorescent photons at high energies, which can impact the measured spectra and reconstructed images. 

\subsubsection{Transmission and secondary photons}
To study the transmission effect and secondary photons, three simulation 
runs with different mass model configurations were performed.   
Apart from a CdTe detector, the first run included the front grid, the 
second run the rear grid and the third run both the front and the rear 
grids of the sub-collimator \#20. The other parts in the STIX whole 
instrument model were not included in order to speed up simulations.

X-ray photons were generated on  a  plane in front of the front grid, 
with directions parallel to the optical axis of STIX. 
Their energies were sampled from a flat spectrum from 3 to 150 keV. 
The positions, energies and of the initial and those reaching the 
front surface of CdTe, as well as flags indicating whether they 
pass through any front (or rear) grid strips or slits, 
were recorded in the simulation outputs. 

\begin{figure}
        \centering
          \includegraphics[width=0.7\linewidth]{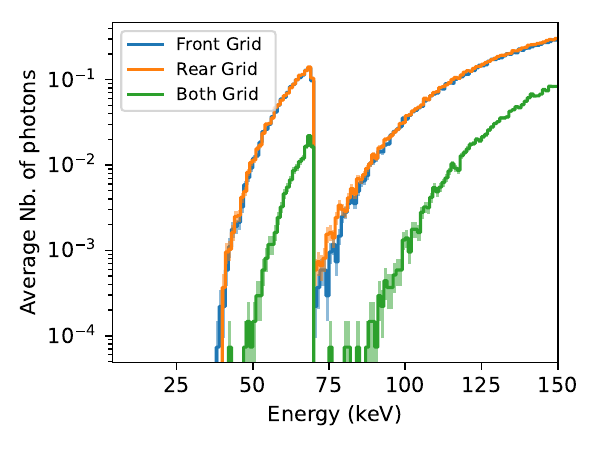}
              \caption{
         Simulated average numbers of photons being detected by a CdTe detector after a photon passes through a 0.4 mm Tungsten strip in the front grid, the rear grid, or passes through both grids as a function of its initial energy. In the case of a single layer of strip, the average number of detected photons per incidence photon is approximately 13.7\% at 69.5 keV,  and 30\% at 150 keV. The directions of photons were parallel to the STIX optical axis.  }
    \label{fig:grid_trans}
\end{figure}

Since we are only interested in interactions with the tungsten grids,  
photons passing through the slits
were discarded during the data analysis. 
Fig.~\ref{fig:grid_trans} shows the simulated average numbers of photons being recorded on the CdTe detector 
after a photon passes through a single layer strip (in the front or rear grid), or two layers of strips.  
The abrupt decrease at 69.5 keV is due to the K-edge of tungsten.  
The differences at energies above the K-edge are attributed to the 
greater detection efficiency of the fluorescent photons from the rear grid as it is closer to the CdTe detector, which are also recorded in the simulation outputs. In the case of a single layer of strip, the average number of detected photons per incidence photon is approximately 13.7\% at 69.5 keV,  and 30\% at 150 keV. 

\begin{figure}
        \centering
   \includegraphics[width=0.45\linewidth]{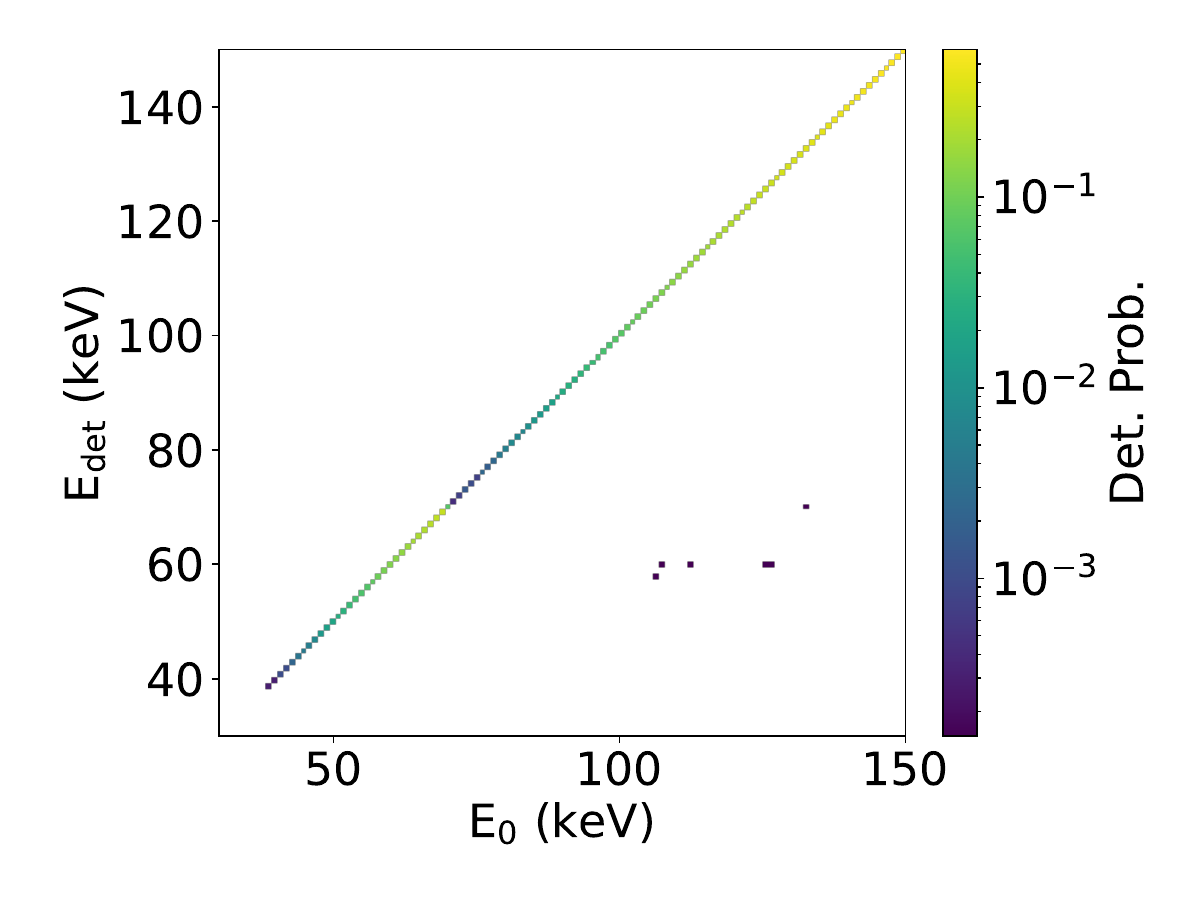}
   \includegraphics[width=0.45\linewidth]{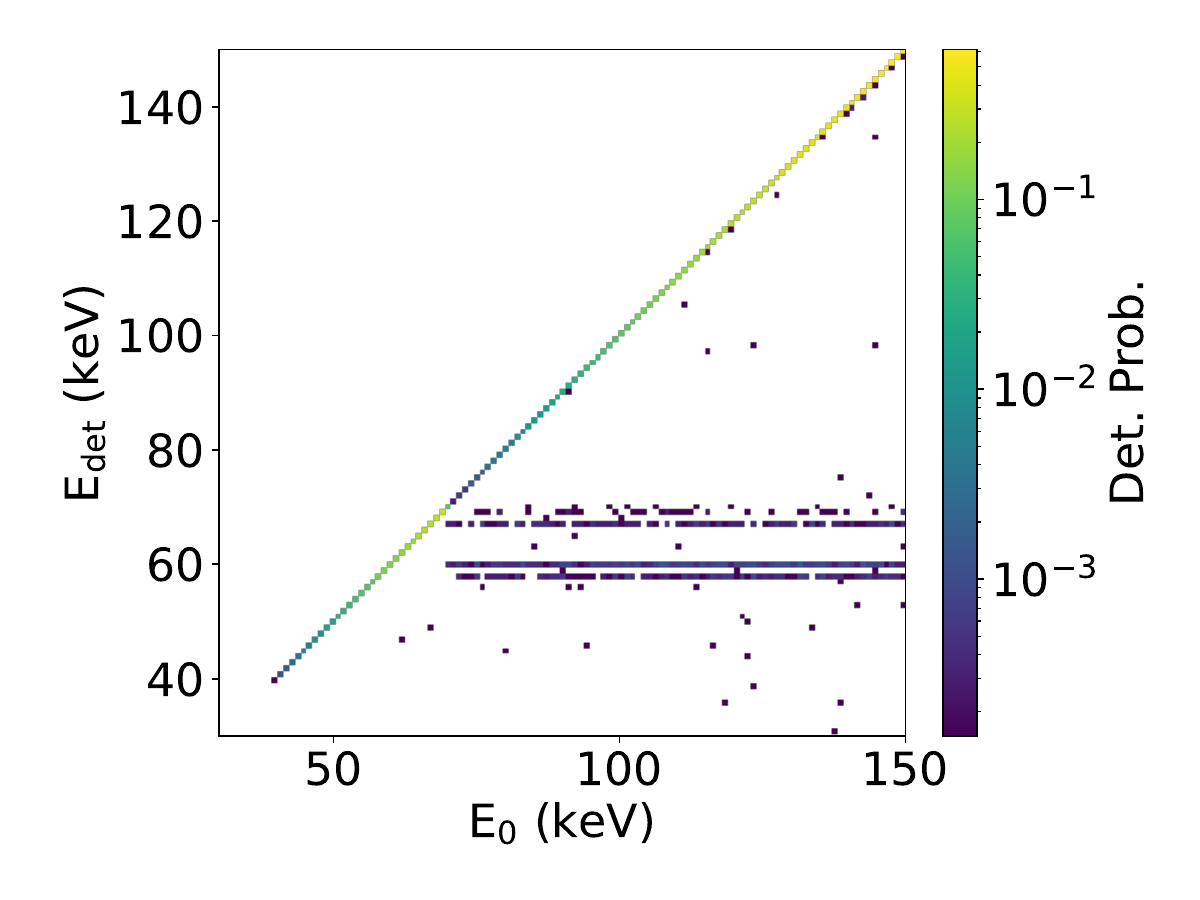}
 \caption{Energies of photons reaching the CdTe detector vs. their initial energies when photons interact with a 0.4 mm strip in the STIX front window (left) and the rear grid window (right). The color scales represent the probabilities. The diagonal lines represent the scenarios where energies of photons are not changed after interactions.
 The horizontal lines from 57 to 70 keV 
 shown in the right panel are attributed to the tungsten emission lines. }
       \label{fig:grid_trans_resp}
\end{figure}

Fig.~\ref{fig:grid_trans_resp} shows the simulated energies of photons reaching the CdTe detector $E_{\rm det}$ vs. their initial energies $E_0$, when photons interact with a strip within the STIX front (left)  or rear (right) 
grid window.
The color scales represent the probabilities.  
The diagonal lines showcase the scenarios 
where the photons have no change of their initial energies 
after interactions. 
The horizontal lines from 57 to 70 keV in the right panel are attributed to the tungsten emission lines. 
The probability of detecting a fluorescent photon when a 70 -- 150 keV photon  interacts with the grid ranges from 0.1\% to 0.2\%.

Emission lines and the extra transmitted photons bring excess counts in the energy range from 50 to 70 keV in STIX measured spectra (see, e.g., \cite{natasha}), they need to be subtracted during spectral analysis.

\subsubsection{Impact of transmitted photons on imaging}
\begin{figure}[htb]
        \centering
          \includegraphics[width=0.7\linewidth]{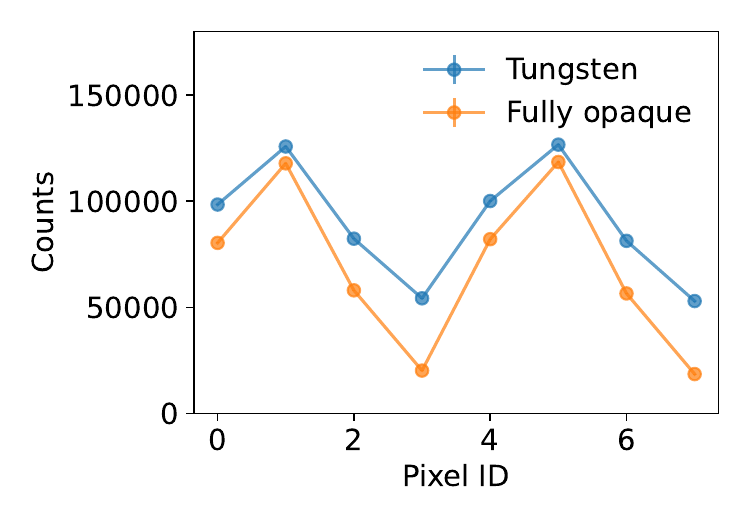}
              \caption{ Comparison of counts detected by CdTe big pixels with 0.4 mm thick tungsten grid strips versus completely opaque grid strips for a 69 keV point-source at infinity. The geometry used in the simulations includes a CdTe detector, the front, and rear grids within sub-collimator \#20, with photons uniformly generated in front of the front grid along the STIX optical axis. 
              }
    \label{fig:grid_pattern}
    %/data2/g4work/g4stix/grid_trans/plot_20240328.jpynotebook
\end{figure}

For an X-ray source, each of the STIX imaging sub-collimators measures a Fourier component, which can be derived from the pattern of the pixel counts \citep{stix2020}. The presence of photons that directly traverse tungsten grids can influence the counts, particularly in the two STIX's nominal energy bins of 63 -- 70 keV and 120 -- 150 keV. 
To study this effect, two simulation runs were conducted.  The sub-collimator \#20, which contains a CdTe detector, the front, and rear grids, was included in the mass models.
In the simulations, primary photons with an energy of 69 keV   are uniformly generated in front of the front grid along the STIX optical axis, simulating the scenario for photons originating from 
a point-source at infinity. In the first run,  photons were tracked normally; whereas in the second run, photons were killed when interacting with tungsten grids.
Thus, tungsten grid strips can be considered completely opaque in the second run. 
A comparison of number of photons reaching the CdTe big pixels for the two runs is shown in  Fig.~\ref{fig:grid_pattern}. 
It can be seen that the amplitude decreases due to the photon passing through the tungsten, which will result in a widening of the reconstructed sources if no correction is applied.

\subsection{Attenuator}
A mechanical attenuator, consisting of six  0.6 mm 
thick L-shape aluminum alloy (type: EN AW-7075) blades,  is inserted automatically  in front of CdTe detectors to reduce  X-ray fluxes when the total trigger rate exceeds a value that a X1 flare can create at 1 A.U.  \citep{stix2020}. When passing through the attenuator, X-rays can be absorbed, scattered, or create secondary 
 photons or electrons. Since the attenuator blades are only about 2 cm to the CdTe detectors,  secondary photons (or electrons) are more likely recorded by those from the tungsten grids. 

To study those effects, we conducted a simulation run with a simplified mass model, which includes the attenuator blades and 32 CdTe detectors. The primary photons had a flat spectrum  from 4 keV to 150 keV and directions parallel to the optical axis. 
The locations, directions, and energies of photons (or electrons) reaching the surface of CdTe were recorded. Then they were killed to exclude the impacts of any physical processes in the CdTe detectors.

\begin{figure*}
        \centering
        \includegraphics[width=0.45\linewidth]{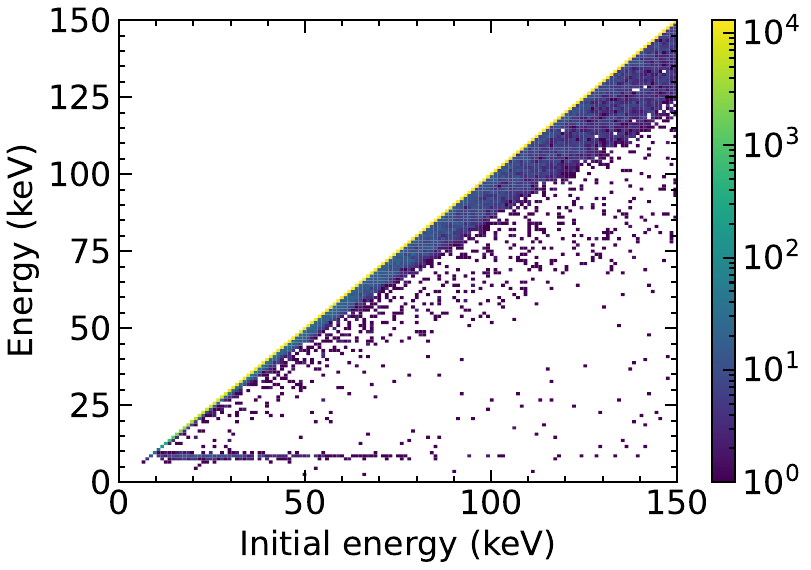}
        \includegraphics[width=0.45\linewidth]{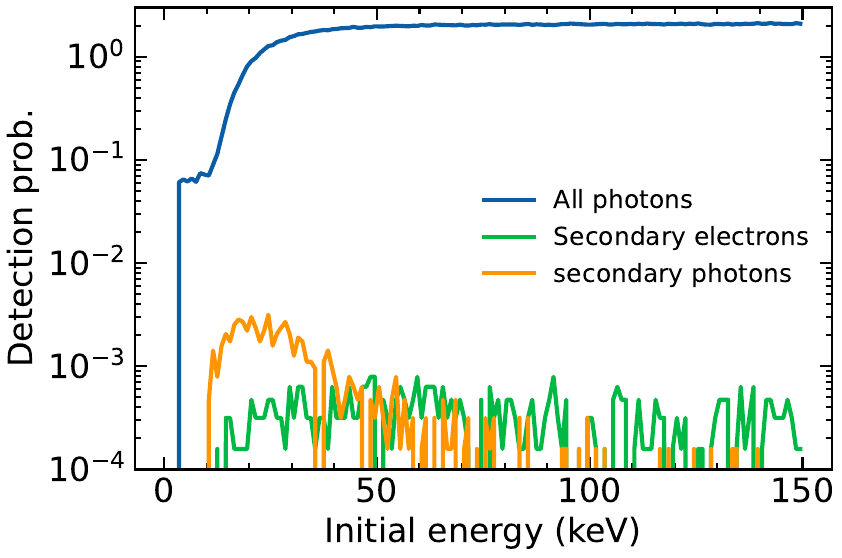}
          \caption{Scatter plot of energies of photons versus CdTe detector energies with respect to initial photon energies entering the attenuator (left), and probability of detecting a photon or secondary particles after a photon passes through the attenuator (right). Note that Compton scattered photons are not considered as secondary particles. Counts recorded by the coarse flare locator (CFL), which is not covered by the attenuator even when it is inserted, are excluded. }
    \label{fig:alures}
\end{figure*}

The left panel of Fig.~\ref{fig:alures} shows a scatter plot  of  the energies of the photons recorded by CdTe detectors with respect to their initial energies.
Copper and zinc are the two main elements apart
from aluminum in the alloy used to manufacture the attenuator.
The excess of recorded counts at around 8.5 keV is due to their characteristic X-rays of  at 8.05 and 8.62 keV, respectively. 

The ratio of the total number of photons  in the non-diagonal
elements is shown in the right panel of Fig.~\ref{fig:alures}. 
As can be seen in the right panel, secondary particles and scattered photons account for approximately 1\% to 13\% of the total detected photons at energies below 20 keV. The excesses at higher energies, which lie below the non-diagonal elements, are mostly due to the Compton-scattered photons, accounting for about 1\% of the recorded photons at a given energy. 
The results are consistent with theoretical predictions: X-rays in the range of a  few keV to hundreds of keV penetrate the attenuator or interact with it
through the photoelectric effect or Compton effect. 
The photoelectric effect is the primary process at energies below 20 keV, 
while Compton scattering becomes dominant at higher energies. 
They can result in the generation of fluorescent X-rays or
Auger electrons (most of which are unable to escape from the material) if they create vacancies in atomic shells.

 \begin{figure}
        \centering
        \includegraphics[width=0.6\linewidth]{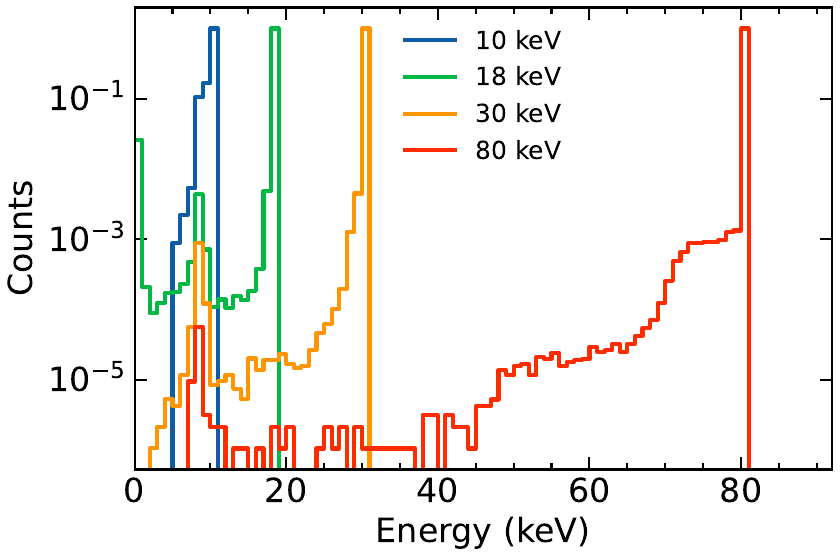}
        \label{fig:atteff}
        \caption{       
         Energy spectra of photons before they reach the CdTe detectors, with initial photon energies of 10, 30, and 80 keV (right). 22\%, 3.6\%, 0.7\%, and 0.9\% of the photons have lower energies than their initial energies, respectively. }
    \label{fig:cdte}
\end{figure}
 \begin{figure}
        \centering
        \includegraphics[width=0.6\linewidth]{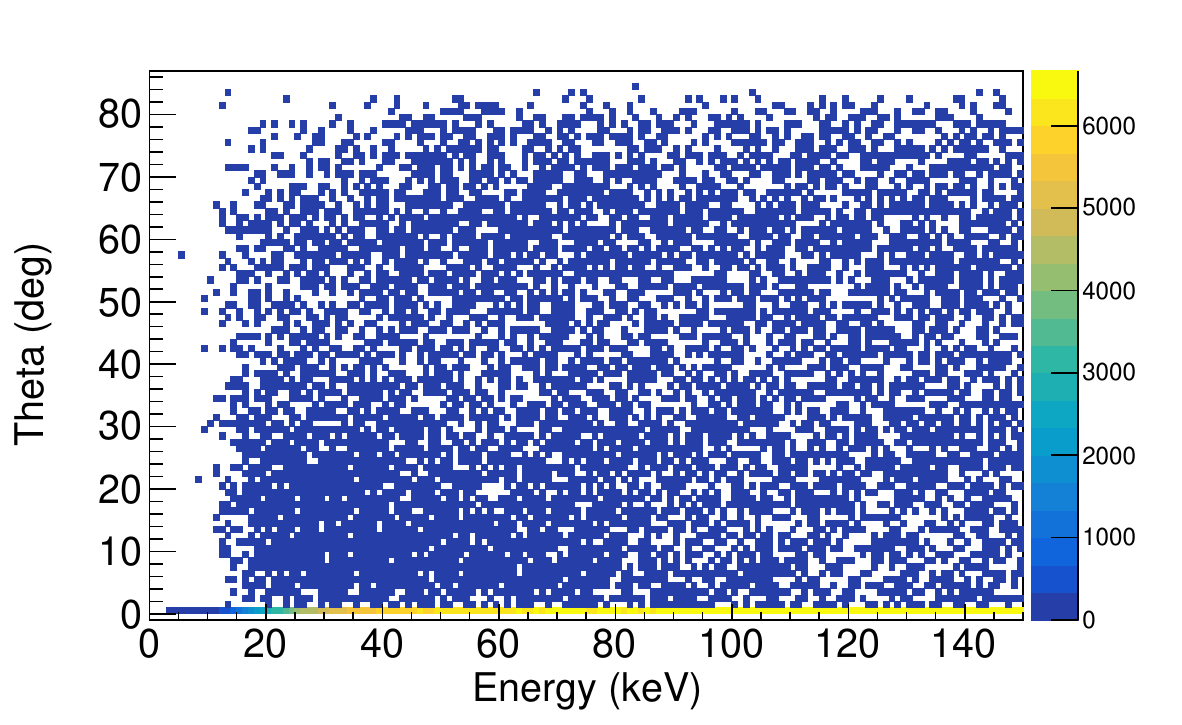}
        \label{fig:atteff}
        \caption{Histogram of the angle of incidence of the recorded photons, plotted against the initial energy of photons. At 10, 30, and 80 keV, the ratios of photons with angle of incidence greater than 0 are 41.0\%, 2.2\%, and 1.3\%, respectively, which are due to Compton scattered photons or fluorescent X-rays.}
    \label{fig:cdte}
\end{figure}
When fluorescence photons and Compton scattered photons are counted,
CdTe detectors record 2-5\% more counts.

\subsection{Effects in the Caliste-SO}
 \begin{figure}
        \centering
        \includegraphics[width=0.7\linewidth]{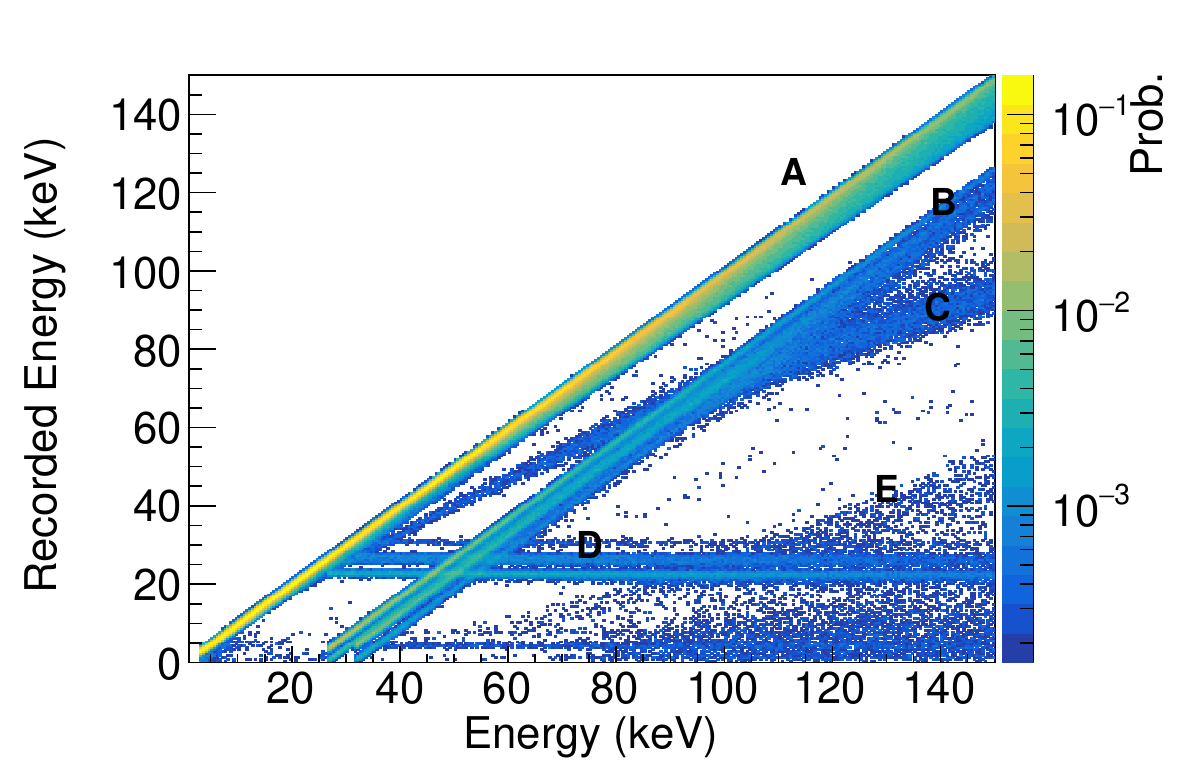}
        \caption{Scatter plot of simulated recorded visible energies in CdTe with respect to  energies of primary photons. Primary photons were shot perpendicular to the surface of a CdTe detector.
        The primary photon energies were sampled from a flat spectrum ranging from 3 to 150 keV. The mass model of Caliste-SO was only included. 
        The distinct clusters annotated with different letters are  caused by different physical processes.
        }
        \label{fig:caliste-effects}
\end{figure}
The STIX instrument utilizes 32 Caliste-SO detectors for X-ray detection \cite{calisteso}. Each Caliste-SO consists of a 1 mm pixelated CdTe sensor and a base containing an ASIC encapsulated in the epoxy resin.

The interaction of X-rays with CdTe sensors involves a complex relationship between X-ray energy and various phenomena. Depending on the energy, physical processes such as photoelectric absorption, Compton scattering, and Rayleigh scattering take place. X-rays that penetrate the CdTe may undergo back scattering or produce fluorescence photons with the surrounding materials, which can subsequently interact with the CdTe. 
Furthermore, the generated charge is 
only partially collected and may be 
smeared out as detailed in Section \ref{sec:effects}.  

To investigate these intricate effects, a simulation run was performed with a
single Caliste-SO in the mass model for a flat spectrum X-ray source ranging from 3 to 150 keV. Photons were incident perpendicular to the CdTe surface. 
This simulation includes all the aforementioned physical processes within the Caliste-SO 
and the detector effects described in Section
\ref{sec:effects}.
Fig.~\ref{fig:caliste-effects} shows the scatter plot of simulated recorded visible energies in all 12 pixels in the Caliste-SO with respect to the energies of primary photons. The colors represent the probabilities.
The distinct clusters labeled with different letters are attributed to various physical processes:
\begin{itemize}
    \item A: Incident photons are fully absorbed by CdTe;  As a result of detector response, the recorded energies are broadened.
    \item B: Fluorescent photons are generated after incoming photons are absorbed by Cd and Te atoms, and the fluorescence photon can escape from the CdTe.
    \item C represents Compton edges. An incident photon can undergo Compton scattering in the CdTe and therefore deposits part of its energy in the CdTe.
    \item D is formed by fluorescent photons of the elements Cd and Te; 
    \item E is attributed to back-scattered photons. Photons that undergo Compton scattering in surrounding materials can be detected by CdTe.
\end{itemize}
\section{Full instrument responses to X-rays}
 \begin{figure}
        \centering
        \includegraphics[width=0.7\linewidth]{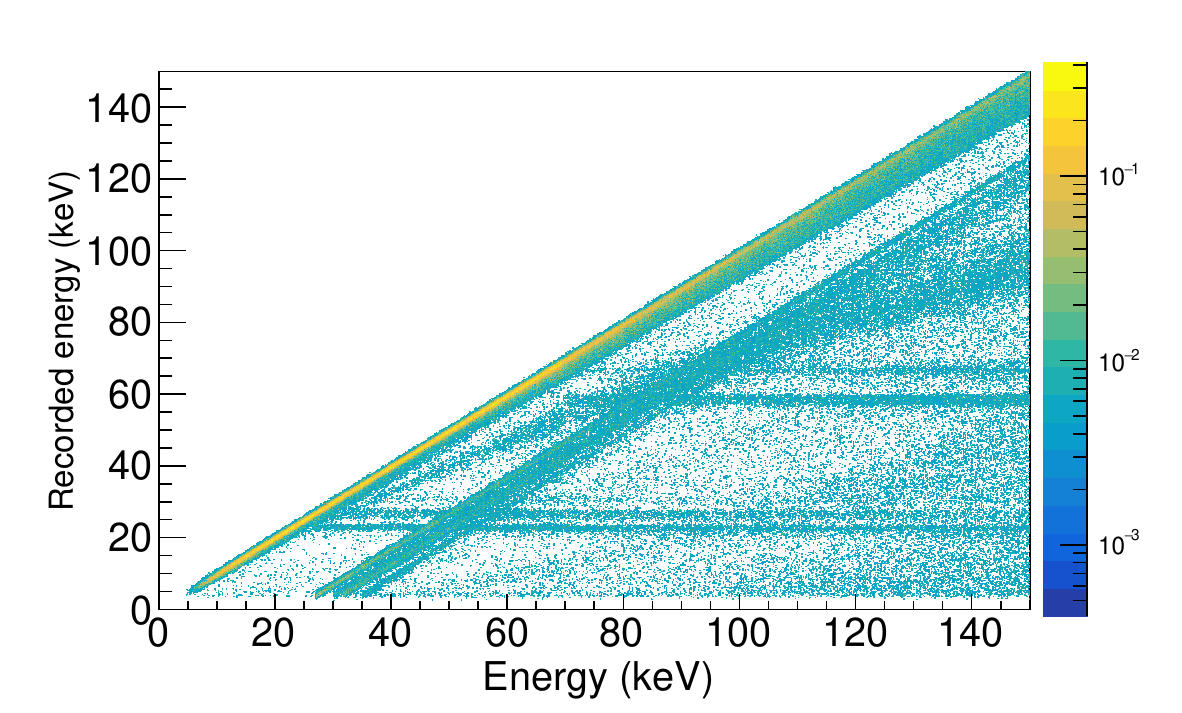}
        \label{fig:fulldet}
        \caption{Response matrix: scatter plot showing simulated recorded visible energies in CdTe relative to energies of primary photons shot in front of the X-ray entrance window. All effects were included in the simulations and the attenuator was not inserted. 
        }
\end{figure}
In order to obtain the response matrices, we perform a simulation run 
with photons distributed across the full energy range of STIX (3--150 keV) and uniformly distributed across the X-ray entrance window. The resulting response matrices provide
 the probability of detecting a photon at a given measured energy for each incident photon energy.
The response matrices generated from our simulations 
represent the instrument's energy redistribution function, accounting for all physical processes discussed in the preceding sections. These matrices are essential for spectral analysis, enabling the conversion of measured count rates to incident photon spectra. The response matrices include the effects of grid transmission and fluorescence,
 detector charge collection efficiency, and energy resolution smearing.

\section{Discussions}
The development and validation of the Geant4-based STIX response model represents a 
significant advance in understanding the instrument's detailed behavior. 
The simulations reveal several important aspects of STIX observations:

\paragraph{Importance of accurate response matrices.}
The detector response matrix (DRM) is fundamental to spectral analysis. Without proper accounting for the instrument's response, derived physical parameters such as plasma temperatures, emission measures, and elemental abundances will be biased. Our simulations demonstrate that including realistic effects such as grid transmission and fluorescence can modify the effective DRM by tens of percent at certain energies, particularly near characteristic X-ray lines and above the tungsten K-edge.

\paragraph{Limitations and uncertainties.}
While our model successfully reproduces the observed Crab Nebula spectrum, several limitations remain. The mass fractions and composition of certain components are based on nominal specifications; actual instrument properties may differ slightly due to manufacturing tolerances and space weathering. Additionally, detector degradation over the mission lifetime introduces time-dependent variations in response that are not fully captured in static response matrices. We recommend generating response matrices at regular intervals based on in-flight calibration data to track these changes.

\paragraph{Future improvements.}
For enhanced spectral analysis, response matrices should ideally be generated individually for each observed flare, incorporating the specific detector conditions (high voltage settings, temperature, gain calibration status) at the time of observation. This approach would provide more accurate uncertainties in derived parameters. Additionally, more comprehensive cross-calibration with other hard X-ray instruments (such as NuSTAR and Fermi-GBM during favorable geometric configurations) would improve the absolute calibration of STIX in its operational energy range.

\section{Conclusion}
We have presented a comprehensive Monte Carlo model of the STIX instrument based on Geant4, which successfully simulates X-ray interactions with all major instrument components. The model incorporates detailed physics including photoelectric absorption, Compton scattering, Rayleigh scattering, and characteristic X-ray fluorescence, along with realistic detector response effects such as charge collection efficiency and energy resolution smearing.

Validation against Crab Nebula observations demonstrates excellent agreement between simulated and observed spectra, confirming the accuracy of our model. We have studied the impacts of key instrumental effects including tungsten grid transmission and secondary photons, attenuator interactions, and detector-specific response features.

The response matrices generated from these simulations are essential tools for STIX data analysis and spectroscopy. They enable accurate conversion of raw count rates to incident photon spectra, accounting for all relevant physical processes. The model and response matrices presented here provide a robust foundation for scientific investigations of solar flare X-ray emissions with STIX and will support the analysis of the entire Solar Orbiter mission dataset.

\bibliographystyle{aa}
\bibliography{citations}
\end{document}